\documentclass{emulateapj}
\usepackage{graphicx}
\usepackage{epstopdf}

\def\be{\begin{equation}}
\def\ee{\end{equation}}
\def\bea{\begin{eqnarray}}
\def\eea{\end{eqnarray}}
\newcommand{\eg}{{\it e.g., }} 
\DeclareGraphicsExtensions{.pdf,.png,.jpg,.eps,.ps}

\journalinfo{Astrophysical Journal Letters, in press}
\slugcomment{}

\shorttitle{The curious case of Lyman Alpha Emitters}
\shortauthors{Acquaviva et al.}

\begin{document}

\title{The Curious Case of Lyman Alpha Emitters: \\ 
Growing Younger from \lowercase{$z$} $\sim$ 3 to \lowercase{$z$} $\sim$ 2?}

\author{Viviana Acquaviva\altaffilmark{1}, Carlos Vargas\altaffilmark{1}, Eric Gawiser\altaffilmark{1}, Lucia Guaita\altaffilmark{2}}

\altaffiltext{1}{Department of Physics and Astronomy, Rutgers, The State University of New Jersey, Piscataway, NJ 08854}
     \altaffiltext{2}{Institutionen f{\"o}r Astronomi, Stockholms Universitet, SE-106 91 Stockholm, Sweden}
    
\begin{abstract}
Lyman Alpha Emitting (LAE) galaxies are thought to be progenitors of present-day L$^*$ galaxies. Clustering analyses have suggested that LAEs at $z \sim 3 $ might evolve into LAEs at $z \sim 2$, but it is unclear whether the physical nature of these galaxies is compatible with this hypothesis. Several groups have investigated the properties of LAEs using spectral energy distribution (SED) fitting, but direct comparison of their results is complicated by inconsistencies in the treatment of the data and in the assumptions made in modeling the stellar populations, which are degenerate with the effects of galaxy evolution. By using the same data analysis pipeline and SED fitting software on two stacked samples of LAEs at $z = 3.1$ and $z = 2.1$, and by eliminating several systematic uncertainties that might cause a discrepancy, we determine that the physical properties of these two samples of galaxies are dramatically different. LAEs at $z = 3.1$ are found to be old (age $\sim$ 1 Gyr) and metal-poor ($Z < 0.2 Z_\odot$), while LAEs at $z = 2.1$ appear to be young (age $\sim$ 50 Myr) and metal-rich ($Z > Z_\odot$). The difference in the observed stellar ages makes it very unlikely that $z$=3.1 LAEs evolve directly into $z$=2.1 LAEs. Larger samples of galaxies, studies of individual objects and spectroscopic measurements of metallicity at these redshifts are needed to confirm this picture, which is difficult to reconcile with the effects of 1 Gyr of cosmological evolution.  

\end{abstract}

\section{Introduction}
Lyman Alpha Emitting (LAE) galaxies have been shown to be building blocks of Milky-Way type galaxies today (\citealt{Gawiser:2007}, \citealt{2010ApJ...714..255G}). The brightness of the Lyman Alpha line allows one to detect these galaxies using the narrow-band technique even when the continuum is faint. As a result, large samples of LAEs have been studied using SED fitting at a variety of redshifts (\eg \citealt{Gawiser:2006a}, \citealt{Nilsson:2007}, \citealt{Gawiser:2007}, \citealt{2007ApJ...655..704L}, \citealt{Pirzkal:2007}, \citealt{2007ApJ...660.1023F}, \citealt{Lai:2007bh}, \citealt{Pentericci:2009}, \citealt{Ono2010}, \citealt{Ono2010b}, \citealt{Ouchi:2010}, \citealt{2011A&A...529A...9N}, \citealt{2011ApJ...733..114G}, \citealt{2011ApJ...733..117F}). The picture emerging from these collective studies is far from homogeneous. While LAEs were once thought to be young, dust-free galaxies experiencing their first episode of star formation, many of these investigations suggest a large spread in their physical properties, and find evidence of an older, more evolved, and dustier stellar population component. 

An important caveat in the interpretation of results comes from the systematic uncertainties introduced by inconsistent modeling of the stellar populations among different groups. This issue was studied in detail by \cite{2011ApJ...737...47A}, where we showed that the use of BC03 (\citealt{2003MNRAS.344.1000B}) and CB11 (\citealt{BCprivatecomm}) stellar population synthesis (SPS) templates, at Solar or variable metallicity, and with or without including nebular emission, gave rise to a scatter in the estimate of age and mass significantly larger than the statistical uncertainty for the same data. 
Further scatter might be created by the use of a different initial mass function (IMF) or of inconsistent statistical estimators of the physical properties of galaxies, such as the best-fit parameters as opposed to the mean of the probability distribution employed by Bayesian statistics.

\section{Objectives, methodology, and data}

\begin{figure}
\begin{centering}
\includegraphics[width=0.8\linewidth]{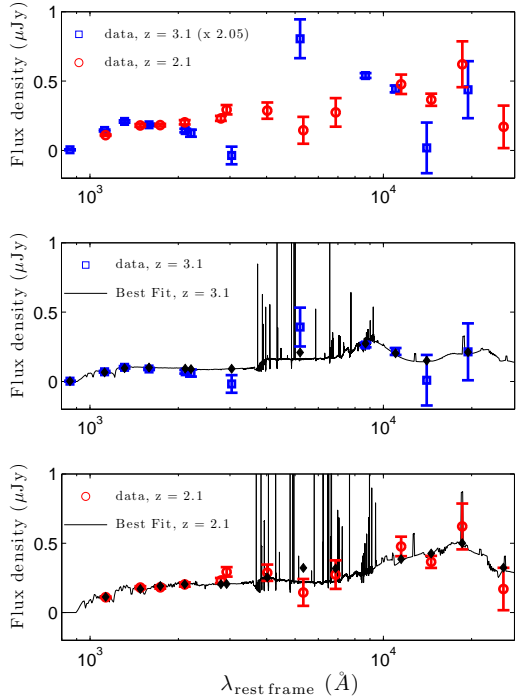}
\caption {Data points and best fit spectra for the full LAE samples. The black diamonds show the flux predicted by the best-fit model in each band. The Ly$-\alpha$ line has been subtracted from the photometry and is not included in the templates used to fit the SED. In the top panel, data at $z = 3.1$ have been offset to account for the different luminosity distance. The dip in the SED at rest-frame wavelengths $\sim$ 1 $\mu$m is typical of low-metallicity, old stellar populations.}
\label{fig:Data}
\end{centering}
\end{figure}

This paper focuses on one particular puzzle, the ages of Lyman Alpha Emitters, and one particular redshift range, the Gyr between $z \sim 3$ and $z \sim 2$. LAEs have been mostly studied through stacking analyses, because their faint continua make it hard to fit the broad-band photometry of each galaxy, although \cite{Ono2010} and \cite{2011A&A...529A...9N} have also attempted to fit the SEDs of bright individual objects at these redshifts. At $z \sim 3$, \cite{Gawiser:2007} and \cite{Lai:2007bh} (hereafter L08) examined the SEDs of average-stacked LAEs, segregated according to their detection in the IRAC 3.6 $\mu$m band, and found that both samples were essentially dust-free, and that LAEs Undetected in IRAC were fairly young (age $\sim$ 0.15 Gyr) and LAEs Detected in IRAC were old (age $\sim$ 1.6 Gyr). At the same redshift, \cite{Nilsson:2007} also found an old median age of 0.85 Gyr, while \cite{Ono2010} found LAEs to be typically young (age $<$ 0.1 Gyr), although a few of the individually fitted K-detected objects appeared to be older. At redshift $z \sim 2$, \cite{2011ApJ...733..114G} revealed the LAEs to be young (age $<$ 0.1 Gyr), and did not find a noticeable difference in the age of LAEs when classified and stacked on the basis of their IRAC 3.6 $\mu$ m band luminosity. \cite{2011A&A...529A...9N} reported finding a young (age $<$ 0.05 Gyr) and an old (age $\sim$ 1 Gyr) stellar population, of comparable mass, in $z = 2.3$ LAEs, although the authors did not report whether the two-component fit was statistically favored with respect to a fit with a single stellar population. 

The purpose of our work is to compare the properties of LAEs at $z\sim~3$ and $z\sim~2$ by using exactly the same pipeline to compose the stacked samples and employing the same, state-of-the-art algorithm to analyze the SEDs. This procedure will discriminate between physical evolution of LAEs and artificial differences introduced by inconsistencies in building the stacks and modeling the stellar populations.  
 
The LAE stacks at $z = 2.1$ are those presented in  \cite{2011ApJ...733..114G}, hereafter Gu11. We consider the full sample, which includes 216 objects, as well as the IRAC-bright and IRAC-dim subsamples, which were built using flux cuts corresponding to the IRAC detection criterion used by L08, appropriately rescaled to account for the difference in luminosity distance. These samples were composed by median stacking, which is less sensitive to outliers than average stacking. For this reason, we also revisit the Detected and Undetected stacks of L08 and use median stacking, and we compose a new median-stacked SED (the full sample at $z$ = 3.1) that comprises 70 LAEs from \cite{Gronwall:2007qd}. At both redshifts, LAEs are selected based on rest-frame Ly$\alpha$ equivalent width $>$ 20 and on the flux in the Ly-$\alpha$ line. The flux limits used in the two surveys lead to very similar rest-frame Ly-$\alpha$ luminosity limits, making the samples at $z = 3.1$ and $z = 2.1$ suitable for a comparative study \citep{2012ApJ...744..110C}.
At $z = 3.1$, we have UBVRIzJK data from MUSYC \citep{Gawiser:2006a} and IRAC photometry, while at $z = 2.1$ we also have H band data. The photometry is computed using IRAF; we use an aperture radius of 1 arcsec in the UV, optical and NIR bands, and 1.25 arcsec in the IRAC bands. The fluxes are aperture-corrected to estimate the total fluxes as described in \cite{Gawiser:2006a}. 
For both samples, objects found in crowded IRAC regions, where imperfect subtraction of bright nearby neighbors would bias the photometry, are discarded before the stacking. 
We cross-checked that the photometric data obtained using SExtractor are compatible with those computed through IRAF. Errors on the stacked photometry are determined by a bootstrap procedure, which accounts for 
both the photometric error and the scatter resulting from the spread in the SEDs of different galaxies. At both redshifts, the latter is the predominant cause of uncertainties; as a result, the size of the error bars and the number of objects in each stack are effectively uncorrelated.

We use the Markov Chain Monte Carlo code GalMC \citep{2011ApJ...737...47A} to fit the spectral energy distribution of these six samples, and the ``GetDist" software from \cite{Lewis:2002ah} to analyze the chains. The data are made available on the MUSYC public data release website at http://physics.rutgers.edu/$\sim$gawiser/MUSYC/data.html. 

\begin{table*}[t!]
  \center
  {\footnotesize
  \resizebox{\textwidth}{!}{
\begin{tabular}{lccccccc}
    \hline
    \hline
    Sample & z & Notes  & Z/Z$_{\odot}$ & Age (Gyr) & E(B-V) & M$^*$ ($10^8$ M/M$_\odot$)& best fit $\chi^2$/d.o f. \\ 
    \hline
    \hline
     full & 3.1 &  & 0.025 [0.005-0.04] & 0.98 [0.84-2.0] & 0.019 [0.-0.022] & 16 [13-19] & 11.9/8 \\ 
    \hline
     full & 3.1& fixed $Z$  & 0.2 & 0.7 [0.6-0.87] & 0.015 [0-0.02] & 13 [12-18] & 14.7/9 \\ 
    \hline
     full & 3.1 & ESF, fixed $Z$  & 0.02 & 0.97 [0.76-2.1] & 0.025 [0-0.031] & 15 [12-18] & 9.5/8 \\ 
    \hline
     full & 3.1 & 2 Pop, fixed $Z$  & 0.02 & 0.45 [0.12-2.1] & 0.0195 [0-0.024] & 15 (37 [0-54]) & 8.1/8 \\ 
    \hline
   full & 2.1 &  & 1.6 [1.4-2.1] & 0.05 [0.02-0.1] & 0.12 [0.1-0.15] & 3.2 [1.9-5.5] &  20.0/9 \\ 
    \hline
     full & 2.1 & fixed $Z$  & 0.2 &0.09 [0.06-0.13] & 0.14 [0.12-0.16] & 6.4 [5.2-7.7] & 33.0/10 \\
    \hline
     full & 2.1 & ESF, fixed $Z$  & 1.0 & 0.07 [0.04-0.11]  & 0.11 [0.09-0.14] & 5.3 [3.7-7.2] & 19.7/9 \\
    \hline
     full & 2.1 & 2 Pop, fixed $Z$  & 1.0 & 0.05 [0.03-0.1]  & 0.12 [0.09-0.14] & 3.7 (0.07 [0-0.2]) & 19.7/9 \\
    \hline
    IRAC Det & 3.1 &  & 0.05 [0.005-0.1]  & 1.0 [0.93 - 2.1] & 0.05 [0-0.06] & 56 [45-69] & 9.7/8  \\ 
    \hline
  IRAC Bright & 2.1 &  & 0.2 [0.03-1.2] & 0.03 [0.01-0.08] & 0.29 [0.27-0.32] & 17 [9-39] & 19.2/9 \\ 
    \hline
   IRAC Und & 3.1 &  & 0.025 [0.005-0.04] & 0.45 [0.4-0.6] & 0.024 [0-0.03] & 6.7 [5-8.5] & 19.5/8 \\ 
    \hline
  IRAC Dim & 2.1 &  & 2.7 [2.3-5.0] & 0.07 [0.02 -0.2] & 0.03 [0 - 0.04] & 1.8 [1.2-2.8] & 22.7/9 \\ 
    \hline
     \hline
  \end{tabular}}
  }
\caption{Mean expectation values and 68\% credible intervals from SED fitting for the six different samples considered in this work. For the two-population fit, we report the mean value of the total mass and the ratio between the mass in the old stellar population and the total mass and its uncertainty in parentheses.}
\label{tab:resSED} 
\end{table*}

\section{Results}

\begin{figure*}[t]
\includegraphics[width=\linewidth]{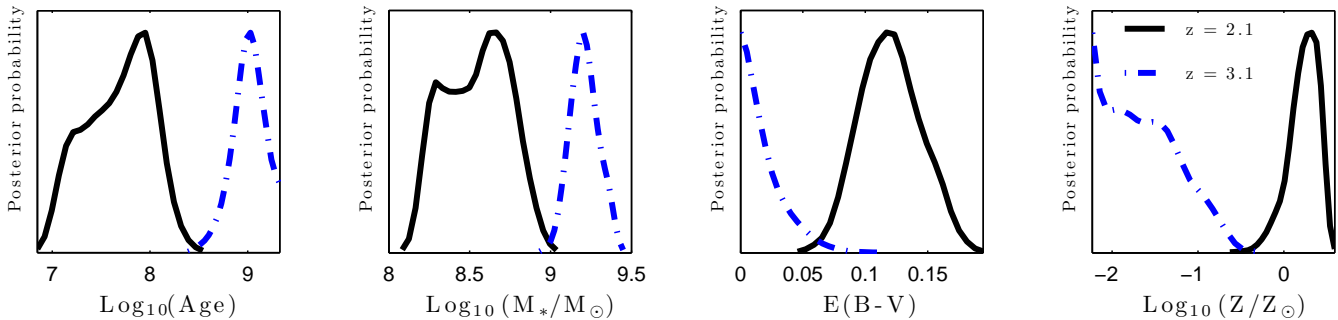}
\caption{Marginalized probability distributions for the SED fitting parameters for $z = 3.1$ (blue, dashed), and $z = 2.1$ (black, solid)  full LAE samples. Curves are normalized to have the same peak height.} 
\label{fig:Prob1D}
\end{figure*}

\begin{figure*}[t]
\includegraphics[width=\linewidth]{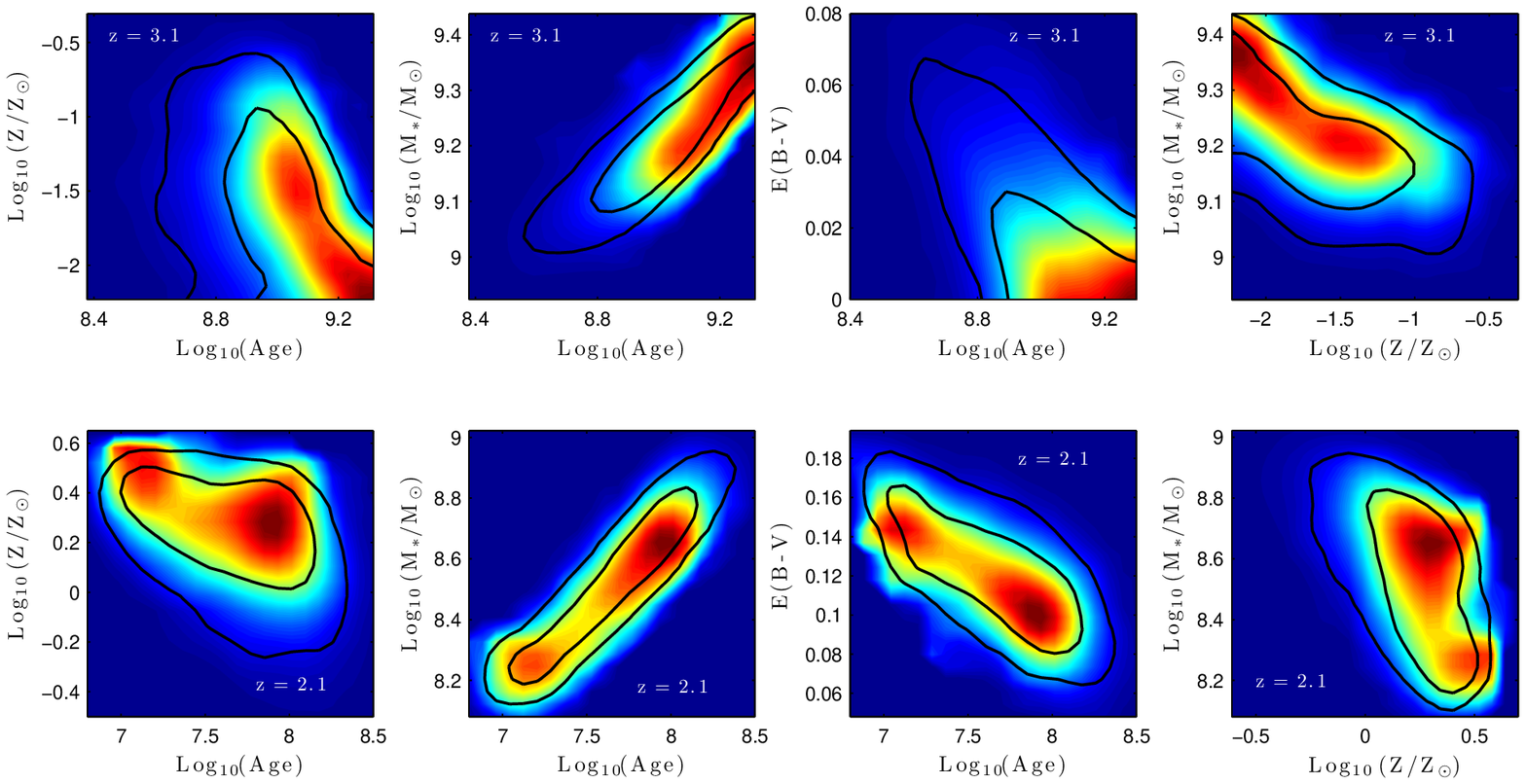}
\caption{2-dimensional marginalized contours illustrating the correlations between the most correlated parameters. The solid lines show the 68 and 95\% error regions for the parameters; the color scale traces the likelihood. For flat priors, lack of exact overlap signals non-Gaussianity of the posterior distribution.}
\label{fig:Prob}
\end{figure*}

The SEDs of the median-stacked full samples at redshifts $z = 3.1$ and $z = 2.1$ are plotted in Fig. \ref{fig:Data}. We begin by fitting four parameters: age since the beginning of star formation (required to be lower than the age of the Universe at that redshift), dust content parametrized by the excess color E(B-V), stellar mass (which is the integral of the star formation rate minus the mass loss due to the life cycle of stars), and metallicity (using a flat prior in log $Z$, as discussed in \citealt{2011ApJ...737...47A}). We use the latest CB11 models, use the Calzetti law \citep{Calzetti2000} to compute the attenuation, assume constant star formation history, favored by the previous analyses of L08 and Gu11, and adopt a Salpeter IMF. We include the contribution of nebular emission as described in \cite{2011ApJ...737...47A}, following the procedure outlined by \cite{2009A&A...502..423S}.
Results from SED fitting are shown in Figs. \ref{fig:Prob1D} and \ref{fig:Prob} and summarized in Table \ref{tab:resSED}. 

Unsurprisingly, given the different shapes of their SEDs, we find strong differences in the properties of $z = 3.1$ and $z = 2.1$ Lyman Alpha Emitters. The higher-redshifts LAEs are dust-free, have strongly sub-Solar metallicities, have mean masses of $\sim 1.5 \times 10^9$ M$_\odot$, and are significantly older, with mean ages of 1 Gyr, than their lower-redshift counterparts, which present a very moderate amount of dust, appear to have Solar metallicity, have mean masses of $\sim 3 \times 10^8$ M$_\odot$, and have mean ages of 45 Myr. 

Results for the  full, IRAC-bright, and IRAC-dim stacked samples of LAEs at z = 2.1 can be compared directly to the ones for the same stacks of Gu11; the main difference is the lower amount of dust (E(B-V) $\sim$ 0.11 vs 0.22 for the  full sample) found by this analysis, which is a direct consequence of the inclusion of nebular emission lines in our code. We can also compare our median-stacked Detected and Undetected samples to the average-stacked Detected and Undetected samples of L08. In this case the main difference is the older age estimated for the Undetected sample, from the 160 Myr of L08 to the 450 Myr found by us. This difference comes in part from the median stacking, which alters the shape of the SED slightly, enhancing the flux in the IRAC 3.6 and 4.5 $\mu$m bands by a factor of 1.8, and in part from allowing metallicity to vary and finding $Z/Z_\odot < 0.2$. The quoted L08 results assumed Solar metallicity, although the authors reported, consistent with our findings, that their age estimate for the Undetected population increased if $Z = 0.2 \,Z_\odot$ was used.

\section{The Mystery of the Ages of LAEs}

The evolutionary picture of LAEs between $z \sim 3$ and $z \sim 2$ suggested by our results is quite puzzling. The same type of galaxies that are found to be old and metal-poor at $z \sim 3$ appear to be young and metal-rich only a Gyr later. Before trying to reflect on the implications of this scenario, we consider several possible artificial sources of such a sharp dichotomy.

\subsection{Data quality tests}
We begin by examining in detail the stacked SEDs of Fig. \ref{fig:Data}. At $z = 3.1$, the best-model has a very reasonable reduced $\chi^2$ of 1.4. This is a desired property because the stacking procedure relies on the strong assumption that ``the typical LAE" exists, and that the artificial SED built by stacking many real SEDs is a reasonable representation of the spectrum of this typical LAE galaxy. A reasonable reduced $\chi^2$ value suggests that the bootstrap uncertainties correctly account for the spread in the SEDs of LAEs and does not illuminate a template incompleteness problem. Nonetheless, there is a relevant amount of scatter in the stacked SED, with three points particularly standing out, the low measurements in the J and IRAC 5.8 $\mu$m band, and the high measurement in the K band. We re-did the SED fitting excluding each of them, and each pair, in turn, and we found that our results are extremely robust to the exclusion of these data points, with changes in the parameter estimates by less than 0.2$\sigma$. We also note that the results at $z = 3.1$ for both age and metallicity are in good agreement with those of \cite{Nilsson:2007}, and that the analysis of \cite{Ono2010} favors low metallicity values. Strongly sub-Solar metallicities were also found in Damped Ly$\alpha$ systems by \cite{2003ApJ...595L...9P}. 

At $z = 2.1$, the SED visually appears to be smoother; however, its reduced $\chi^2$ has a slightly higher value of 2.2. This feature suggests that the stacking at this redshift might be an overly aggressive compression of the information contained in the individual SEDs. Breaking down the sample in IRAC-bright and IRAC-Dim LAEs in the attempt to increase the homogeneity of the stacks does not afford a better fit, although it shows that the properties of these two stacks follow closely the ones of the  full sample. 

\subsection{Metallicity}
The SED fitting parameters are correlated with each other, and in particular, there is a well-know age-metallicity relation that can be seen easily as the axis of the 2-D credible region in the age-metallicity plane of Fig. \ref{fig:Prob}. Our estimates of metallicity depend on the SPS templates we use and (although weakly) on the flat prior on $\log Z$ that we assume. We think that the latter is physically motivated, and we have no reason not to trust the models; however, it is possible that the observed difference in metallicity between LAEs at the two redshift is overestimated. If the true metallicity at $z = 3.1$ were higher, and the true metallicity at $z = 2.1$ were lower, this might help reconcile the gap in the physical properties of these galaxies. To test this hypothesis, we run our SED fitting code on the  full samples at the two redshifts, holding the metallicity fixed at $Z = 0.2 Z_\odot$; this value is suggested by the spectroscopic analysis of three LAEs at $z \sim 2.3$ by \cite{2011ApJ...729..140F}.  Results are shown in Fig. \ref{fig:Syst} and listed in Table \ref{tab:resSED}. As expected, the estimates for the mean ages of the stellar population shift in the desired direction, but remain incompatible at several sigma level. This result seems to exclude metallicity as the primary source of the gap in the age of LAEs, at least in this stacking analysis. 

\subsection{Star Formation History}
\begin{figure*}[t]
\includegraphics[width=\linewidth]{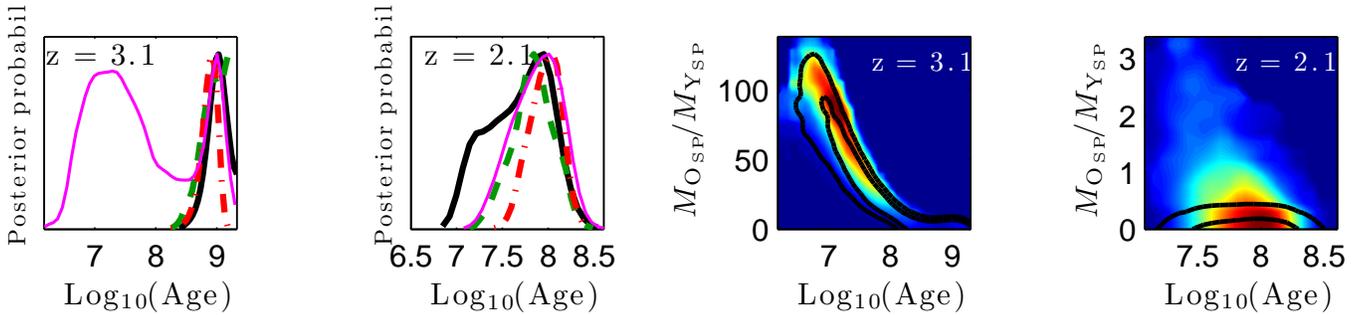}
\caption{{\it Left two panels}: Variation in the posterior probability distribution of age of the full samples, for different assumptions described in the text. The black (solid, thick) line is the reference case ran with CSF and varying metallicity. The red (dotted-dashed) line is for $Z$ fixed at $0.2 Z_\odot$, the green (dashed) line corresponds to exponential star formation, and the magenta (solid, thin) line is the age of the primary population when a second, 1 Gyr old SSP is added. {\it Right panels:} Probability distribution of the fraction of mass in the old stellar population versus the age of the primary population. For the $z$ = 3.1 LAEs, a young stellar component is allowed only when most of the mass is in the old stellar population.}
\label{fig:Syst}
\end{figure*}

The assumption of a particular functional form for the star formation history of LAEs can also influence the determination of ages. We investigate this issue by repeating the SED fitting procedure on the  full samples using exponentially increasing and decreasing star formation histories (SFHs), $\psi(t) \propto e^{\pm t/\tau}$. By sampling in $1/\tau$, both increasing and decreasing SFHs are explored as part of a contiguous parameter space, as explained in \cite{2011ApJ...737...47A}. This parametrization is also able to capture constant or starburst SFHs for appropriate values of $\tau$/age. The results of this SED fitting run (where the metallicities were held fixed at the best-fit value for both samples) are shown in Fig. \ref{fig:Syst}; in both cases the estimates of ages do not change significantly. The quality of the fit does not improve significantly in either case by allowing this larger range of star formation histories, as can be seen in Table \ref{tab:resSED}. The data favor a nearly constant SFH at both redshifts, with a best-fit value of $\tau$ around 4 Gyr for both samples.
Gu11 had also performed the SED fitting using these SFHs for the $z=2.1$ sample, albeit with a different algorithm, and had found similar results.

\subsection{A hidden old stellar population}
A possible explanation of the difference in the physical nature of LAEs at $z = 3.1$ and $z = 2.1$ might be the presence of stellar components of different ages.
We test this hypothesis by performing the SED fitting on the full samples using two stellar populations, fixing the age of the second stellar population at 1 Gyr, and introducing the stellar mass of the second stellar population as a free SED fitting parameter. We show the results of this test for a single-burst model that would solely contain old stars, although we have also considered constant or linearly increasing SFHs for the second stellar population, as suggested by some recent observations (\eg \citealt{2011ApJ...742..108F, 2010PhDT.......132L} and references therein). For $z = 2.1$ LAEs, we also consider a few different values of metallicity and dust content for the second population. At $z = 2.1$, this test is meant to reveal an older stellar population; at $z = 3.1$, it may allow us to single out two components of different ages, although we do not expect to find a large component of very young stars, since they would dominate the rest-frame UV part of the total spectrum. In the case of $z = 3.1$ LAEs, adding a second old stellar population slightly improves the quality of the fit (from $\chi^2_r$ = 1.4 to $\chi^2_r$ = 1.0). Young ages for the primary stellar population are allowed, but only when the fraction of mass in this stellar population is small, as seen in the third panel of Fig. \ref{fig:Syst}. For example, ages of the order of $100$ Myr are allowed at $68\%$ confidence when the mass in young stars is less than $10\%$ of the total stellar mass. As expected since the two stellar populations are so similar, the mass ratio in the two populations is unconstrained.  For $z = 2.1$ LAEs, the quality of the fit does not improve by adding a second stellar population, and again the age estimate for the first stellar population does not change from the value of $\sim$ 50 Gyr favored by the single population fit, and the mass fraction in the ``Old" stellar population is found to be consistent with zero, and $< 50\%$ at $2\sigma$ confidence for all cases considered. We conclude that the two-component scenario cannot be confirmed by analyzing the stacked SEDs, although the improvement in the $\chi^2$ at $z$ = 3.1 might indicate a heterogeneity of the stellar populations, masked by the stacking process.

\section{Conclusions}

The clustering analysis of LAEs \citep{2010ApJ...714..255G} has shown that LAEs at $z \sim 3$ and $z \sim 2$ are hosted by dark matter halos that can be in a direct progenitor-descendant relationship.
 An intriguing possibility is that they might represent two different stages of the same galaxy population. We test this hypothesis by using the same data analysis pipeline and SED fitting software; this strategy allows us to discriminate between evolution in the galaxy properties and the effects of different assumptions in the processing of data and the modeling of the stellar populations.
We find that the physical properties of LAEs at $z = 3.1$ are very different from those of LAEs at $z = 2.1$, which appear to be several times younger and more metal-rich than their higher-redshift counterparts. If these samples are a fair representation of LAEs at each redshift, this result directly rules out the hypothesis that z=3.1 LAEs evolve into z=2.1 LAEs, implying that the LAE phase lasts significantly less than 1 Gyr. Moreover, such stark differences between emission-line selected galaxy samples become difficult to explain as the effect of 1 Gyr of cosmological evolution.

We find hints that stacking analysis, meant to provide insights on the nature of the {\it typical} LAE galaxy, might be overly aggressive and perhaps unable to capture the multi-component nature of LAEs at each redshift. To be properly investigated, these issues require constraints from SED fitting of large samples of individual objects, as well as
spectroscopic constraints on the metallicity of LAEs.  The former have been especially challenging to acquire so far, because of the lack of deep data in the observed NIR region of the spectrum. This gap is being filled by deep YJH HST observations by the Cosmic Assembly Near-infrared Deep Extragalactic Survey (CANDELS, \citealt{2011ApJS..197...35G,2011ApJS..197...36K}), and we plan to address this issue by using CANDELS data in the MUSYC fields in a subsequent paper (Vargas et al 2012, in preparation).

\acknowledgments
It is a pleasure to thank Kamson Lai for advice on the photometry at $z$ = 3.1, the Astrophysics group at Rutgers, in particular Saurabh Jha and Curtis McCully, and the anonymous referee, for useful conversations. This research was supported in part by the American Astronomical Society's Small Research Grant Program.\\

\end{document}